\documentclass{aastex631}
\submitjournal{ApJ}
\usepackage{amssymb}
\usepackage{amsmath}
\usepackage{tabularx}
\usepackage{array,etoolbox}

\usepackage{longtable}
\newcounter{magicrownumbers}
\newcommand\rownumber{\stepcounter{magicrownumbers}\arabic{magicrownumbers}}

\newcommand{\x}{$\times$ }

\received{September 13, 2021}
\revised{April 25, 2022}
\accepted{\today}

\submitjournal{ApJ}

\shorttitle{}
\shortauthors{}

\begin{document}

\title{Vertical distribution of cyclopropenylidene and propadiene in the
  atmosphere of Titan}

\correspondingauthor{Karen Willacy}
\author[0000-0001-6124-5974]{Karen Willacy} 
\affil{Jet Propulsion Laboratory, California Institute of Technology 
4800 Oak Grove Drive, Pasadena, CA 91109, USA}

\author{SiHe Chen}\affiliation{Division of Geological and Planetary Sciences, Caltech,
  Pasadena, CA}

\author[0000-0001-9897-9680]{Danica J. Adams}\affiliation{Division of Geological and Planetary Sciences, Caltech,
  Pasadena, CA}

\author[0000-0002-4263-2562]{Yuk L. Yung}
\affiliation{Division of Geological and Planetary Sciences, Caltech,
  Pasadena, CA}

\begin{abstract}
  
  Titan's atmosphere is a natural laboratory for exploring the
  photochemical synthesis of organic molecules. 
Significant recent advances in the study of the
atmosphere of Titan include: (a) detection of C$_3$ molecules: C$_3$H$_6$,
CH$_2$CCH$_2$, c-C$_3$H$_2$, and (b) retrieval of C$_6$H$_6$, which is formed primarily
via C$_3$ chemistry, from Cassini-UVIS data. The detection of
$c$-C$_3$H$_2$ is of particular significance since ring molecules are
of great astrobiological importance. Using the Caltech/JPL
KINETICS code, along with the best available photochemical rate
coefficients and parameterized vertical transport, we are able to
account for the recent observations. It is significant that ion
chemistry, reminiscent of that in the interstellar medium,  plays a
major role in the production of c-C$_3$H$_2$ above 1000 km. 
\end{abstract}


. 

\section{Introduction} \label{sec:intro}

Titan's atmosphere comprises a rich hydrocarbon chemistry unlike
anything seen elsewhere in the solar system outside of Earth.  The
conditions are likely similar to those found on the early Earth before
the build-up of oxygen.  They also share characteristics with the
experiments of \cite{mu59} who demonstrated that complex organics can
be synthesised by the irradiation of mixtures of simple gases in a
reducing environment.   Titan therefore provides an ideal laboratory
in which the chemistry and synthesis of complex organics and prebiotic
molecules can be explored.

Our knowledge of the composition of Titan's atmosphere has been
greatly enhanced by observations by Cassini and ALMA.   In particular,
several molecules
with three carbon atoms have been observed in Titan's atmosphere,
including CH$_3$C$_2$H, C$_3$H$_6$ \citep{lombardo19a, nixon13, magee09}, and
C$_3$H$_8$ \citep{lombardo19a, nixon13, vinatier10}, and recently two more -- cyclopropenylidene
(c-C$_3$H$_2$) \citep{nixon20} and propadiene (CH$_2$CCH$_2$) \citep{lombardo19b} -- have been added to
the list.  Additionally, ions with $m/z$ corresponding to C$_3$H$_3^+$
and C$_3$H$_5^+$ have been observed by Cassini/INMS \citep{mandt12}.  These observations
provide good constraints on the modeling of the C$_3$H$_n$
species which are important for controling the
abundance of larger molecules such as benzene.

In this paper we focus on the chemistry of C$_3$ hydrocarbons.  
These have 
previously been modeled by \cite{hebrard13}, \cite{li15} and
\cite{vuitton19}.  In the light of the new detection of $c$-C$_3$H$_2$ \citep{nixon20}
we revisit their chemistry and 
update our previous model \citep{li15,willacy16} to include
the isomers of C$_3$H$_2$ and C$_3$H, and 
ion-molecule chemistry. We discuss these updates in Section~\ref{sec:models}. We
present our results and compare them to the observations in
Section~\ref{sec:results}. Finally, we discuss the results and
conclusions in Section~\ref{sec:disc}. 

\section{\label{sec:models}Models}

We use the Caltech/JPL coupled photochemical/transport model, KINETICS
\citep{allen81,yap84}. The model solves the mass continuity equation from
Titan's surface to 1500 km altitude 
\begin{equation}
    \frac{\partial n_i}{\partial t} + \frac{\partial \phi_i}{\partial z} = P_i - L_i
    \end{equation}
where $n_i$ is the number density of species $i$ and $P_i$ and $L_i$
are the production and loss rates respectively.  $\phi_i$ is the vertical flux of species $i$ calculated from
\begin{equation}
    \phi_i = - \frac{\partial n_i}{\partial z} (D_i + K_{zz}) - n_i \left (\frac{D_i}{H_i} + \frac{K_{zz}}{H_\alpha} \right ) - n_i \frac{\partial T}{\partial z} \frac{(1+\alpha_i) D_i + K_{zz}}{T}
    \end{equation}    
where $D_i$ and $H_i$ are the molecular diffusivity coefficient and
the scale height of species $i$, $H_{\alpha}$ is the atmospheric scale
height, $\alpha_i$ is the thermal diffusion coefficient of $i$, $T$ is
the temperature and $K_{zz}$ is the eddy diffusion coefficient. We use
the values of $K_{zz}$ derived by \cite{li14}: 
\begin{equation}
log K_{zz}(z) = \left\{
\begin{array}{ll}
log(3 \times 10^3) , &  z < z_1\\
\\
log(3 \times 10^3) \dfrac{z_2 - z}{z_2 -z_1} + log(2 \times 10^7)
  \dfrac{z - z_1}{z_2 - z_1} , &  z_1 \leq z < z_2  \\
\\
log(2 \times 10^7) \dfrac{z_3 - z}{z_3 - z_1} + log(2 \times 10^6) \dfrac{z - z_2}{z_3 - z_2} , &  z_2 \leq z < z_3\\
\\
log(2 \times 10^6) \dfrac{z_4 - z}{z_4 - z_3} + log(4 \times 10^8) \dfrac{z - z_3}{z_4 - z_3} , &  z_3 \leq z < z_4\\
\\
log(4 \times 10^8) , &  z \geq z_4
\end{array}
\right.
\end{equation}
where $z_1$ = 120 km, $z_2$ = 300 km, $z_3$ = 500 km and $z_4$ = 1000
km.  
The atmospheric density and temperature profiles are also taken from
\cite{li15}, and are based on the T40 Cassini flyby\citep{westlake11}.

 The aerosols in our model both absorb radiation and provide surfaces
 for condensation. Their mixing ratio and surface area as a function
 of altitude are taken from \cite{lavvas10}. The properties were
 derived from a microphysical model validated against Cassini/Descent
 Imager Spectral Radiometer observations. To calculate the absorption
 of UV by dust we assume absorbing aerosols with extinction
 cross-sections that are independent of wavelength \citep{li14,li15}.  
 
The chemical network has been extended to
include the isomers of C$_3$H and C$_3$H$_2$ using the networks of
\cite{hebrard13} and  \cite{loison17}.  Additionally, ion-molecule
chemistry has been added with reactions from \cite{vuitton19} and the
astrochemical databases UMIST12 \citep{rate12}, and KIDA \citep{kida}.
The full list of species is given in Table~\ref{tab:molecules} and the
reaction network can be found in
Appendix~\ref{app:react}.

The
rate coefficients for photodissociation of the isomers of C$_3$H, C$_3$H$_2$ are
taken from \cite{hebrard13}.  Rate coefficients for other photodissociation and
photoionization processes are taken from the standard KINETICS
database.

\subsection{Boundary conditions}
The lower boundary of our model is the surface of Titan and the upper
boundary is at 1500 km.   For H and H$_2$ the flux at the lower
boundary is zero and at the top of the atmosphere these species are
allowed to escape at their Jeans escape velocity of (2.5
$\times$ 10$^4$  cms$^{-1}$ and 6.1 $\times$ 10$^{3}$
cms$^{-1}$ respectively). For all other neutral gaseous species the
concentration gradient at the lower boundary is assumed to be zero,
while they have zero flux at the top boundary.  Ions have a mixing
ratio of zero at $z$ = 0 km, and at the upper boundary their flux is
zero.

The mixing ratio of N$_2$ is held fixed at 0.95 at all altitudes.  
The mixing ratio of CH$_4$ is fixed to the observed (super-saturated) values 
\citep{niemann10} below the tropopause and allowed to vary above this.

\begin{table}
  \caption{\label{tab:molecules}Species included in the model.}
  \begin{tabularx}{\linewidth}{l X}
\hline
Family & Molecule \\
\hline
& H, H$_2$\\
& \\
Hydrocarbons & C  CH  CH$_2$ $^3$CH$_2$  CH$_3$  CH$_4$  C$_2$  C$_2$H  C$_2$H$_2$ C$_2$H$_3$  C$_2$H$_4$  C$_2$H$_5$  C$_2$H$_6$  C$_3$ $l$-C$_3$H  $c$-C$_3$H $l$-C$_3$H$_2$ $c$-C$_3$H$_2$ $t$-C$_3$H$_2$ C$_3$H$_3$   CH$_2$CCH$_2$  CH$_3$C$_2$H  C$_3$H$_5$  C$_3$H$_6$ C$_3$H$_7$  C$_3$H$_8$  C$_4$H  C$_4$H$_2$   C$_4$H$_3$  C$_4$H$_4$  C$_4$H$_5$  1-C$_4$H$_6$  1,2-C$_4$H$_6$  1,3-C$_4$H$_6$  C$_4$H$_8$ C$_4$H$_9$  C$_4$H$_{10}$  C$_5$H$_3$  C$_5$H$_4$  C$_6$H  C$_6$H$_2$  C$_6$H$_3$  C$_6$H$_4$ C$_6$H$_5$ $l$-C$_6$H$_6$  C$_6$H$_6$  C$_8$H$_2$\\
& \\
Nitrogen-molecules & N  N(2D) N$_2$ NH  NH$_2$  NH$_3$  
                     CN  HCN  H$_2$CN  CHCN  HC$_3$N  HC$_5$N  CH$_2$CN  CH$_3$CN
                     C$_2$H$_3$CN  CH$_3$C$_3$N
                     C$_3$N  C$_5$N 
                     C$_2$N$_2$ C$_4$N$_2$ C$_6$N$_2$  \\
& \\
Ions &  H$^+$          H$_2^+$         N$^+$       N$_2^+$   NH$^+$         NH$_2^+$        NH$_3^+$        NH$_4^+$      HN$_2^+$        CN$^+$         HCN$^+$         HCNH$^+$  H$_3^+$         C$^+$          CH$^+$ CH$_2^+$        CH$_3^+$        CH$_4^+$      
                 CH$_5^+$        C$_2^+$         C$_2$H$^+$        C$_2$H$_2^+$       C$_2$H$_3^+$       C$_2$H$_4^+$       C$_2$H$_5^+$       CH$_3$CH$_3^+$     C$_2$H$_7^+$       C$_3^+$       
                 C$_3$H$^+$        c-C$_3$H$_2^+$     $l$-C$_3$H$_2^+$     c-C$_3$H$_3^+$     $l$-C$_3$H$_3^+$     CH$_3$CCH$^+$     C$_3$H$_5^+$       C$_3$H$_6^+$       C$_3$H$_7^+$       C$_4$H$^+$      
                 C$_4$H$_2^+$       C$_4$H$_3^+$       C$_4$H$_4^+$       C$_4$H$_5^+$       C$_4$H$_9^+$          C$_6$H$_3^+$         C$_6$H$_5^+$     
                C$_6$H$_6^+$     C$_6$H$_7^+$      \\
\end{tabularx}

\end{table}

We also include processes removing H from the gas by heterogeneous
reaction with haze particles \citep{sekine08a,sekine08b}.

\section{\label{sec:results}Results}
\subsection{C$_3$H$_2$}
C$_3$H$_2$ can exist in three forms: $c$-C$_3$H$_2$,
(cyclopropenylidene), $l$-C$_3$H$_2$ (propadienylidene) and
$t$-C$_3$H$_2$ (propynylidene). 
The cyclic form, the most abundant isomer, was recently identified for the
first time in Titan's atmosphere by \cite{nixon20}.
The molecule was detected by observations taken in  2016 and 2017 with 
column densities of 3-5 $\times$ 10$^{12}$ cm$^{-2}$ and
1-2 $\times$ 10$^{12}$
cm$^{-2}$ for the two epochs respectively.  $t$-C$_3$H$_2$ and $l$-C$_3$H$_2$ remain
undetected.

\begin{figure}[!h]
    \centering
    \includegraphics[ width=0.5\linewidth]{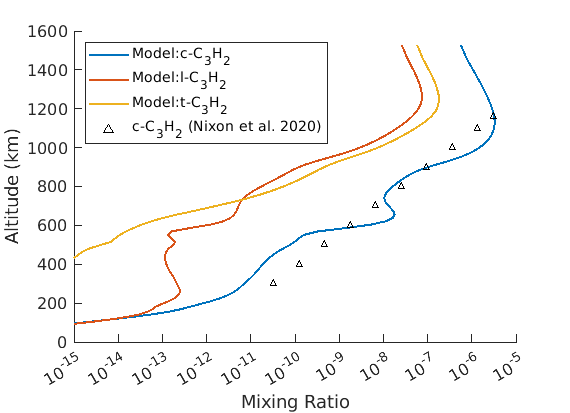}
    \caption{\label{fig:c3h2}Model mixing ratios of $l$-C$_3$H$_2$ (red),
      $c$-C$_3$H$_2$ (blue) and
      $t$-C$_3$H$_2$ (yellow).  Also shown are the observational results
      from \cite{nixon20} assuming a gradient profile for the
      abundance of $c$-C$_3$H$_2$ through the atmosphere above 350 km
      (open triangles). } 
\end{figure}

Our model includes all three structural isomers and their predicted
mixing ratios are shown in Figure~\ref{fig:c3h2}. The mixing ratio
of $c$-C$_3$H$_2$ is in excellent
agreement with the best fit vertical abundance profile derived by
\cite{nixon20}.  The model column density of this
molecule (N($c$-C$_3$H$_2$) = 2.6 \x 10$^{12}$ cm$^{-2}$) is consistent
with that observed.  Similar column densities were predicted by the
models of \cite{vuitton19} (for a single C$_3$H$_2$ isomer) and
\cite{hebrard13} (who considered the three isomers as separate species). 
The predicted column densities of the other two isomers are lower, with
N($l$-C$_3$H$_2$) = 1.6 \x 10$^{11}$ cm$^{-2}$ and N($t$-C$_3$H$_2$) = 6.9
\x 10$^9$ cm$^{-2}$.

\begin{figure}
  \centering
  \includegraphics[width=0.80\linewidth]{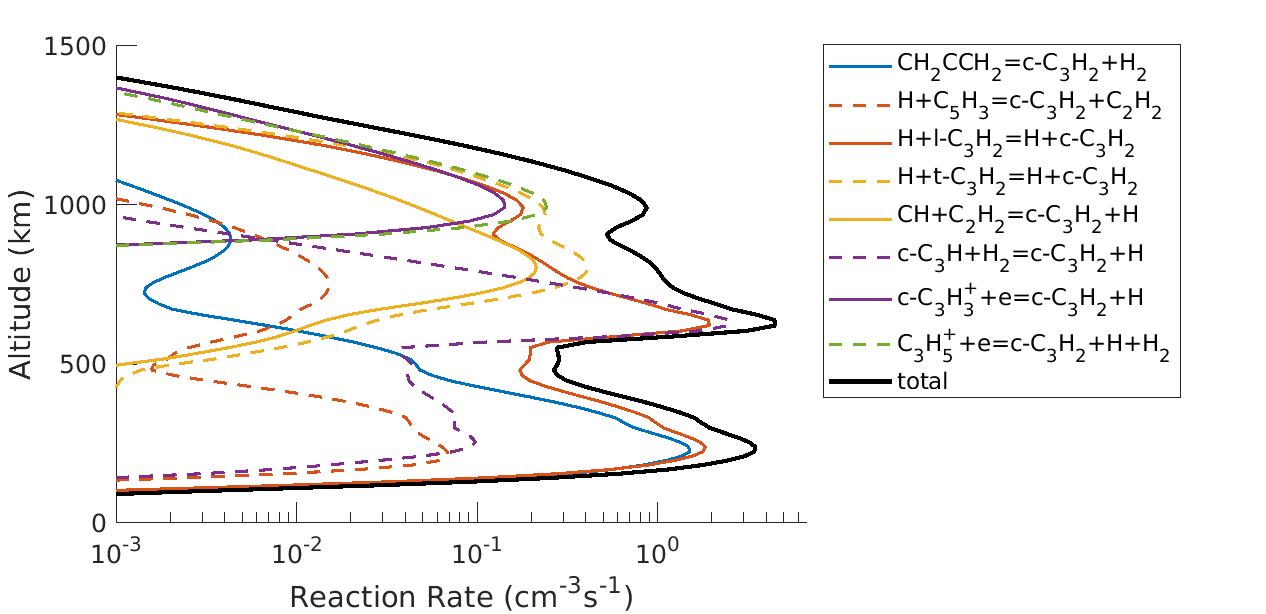}
  \includegraphics[width=0.80\linewidth]{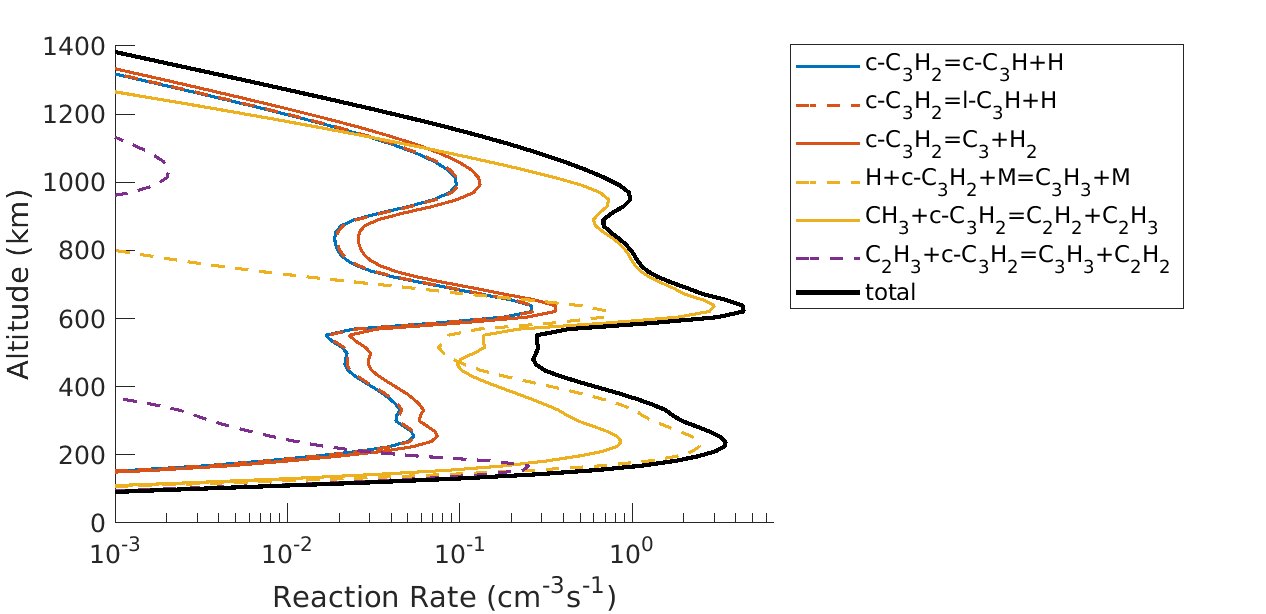}
  \caption{\label{fig:cC3H2_prod}Production (top) and destruction
    (bottom) rates for
    $c$-C$_3$H$_2$.  Destruction by photodissocation produces
    $l$-C$_3$H or $c$-C$_3$H.}
\end{figure}

The main production and loss reactions for $c$-C$_3$H$_2$ and their
rates are shown in Figure~\ref{fig:cC3H2_prod}.
The important reactions vary with
altitude, with electron recombination of $c$-C$_3$H$_3^+$ (R1159) and
C$_3$H$_5^+$ (R1183)
dominating above 1200 km and isomerization of $l$-C$_3$H$_2$ (R22) and
$t$-C$_3$H$_2$ (R23) by reaction with H atoms below this. Photodissication
of CH$_2$CCH$_2$ is important at all altitudes.  The reaction between
CH and C$_2$H$_2$ also forms $c$-C$_3$H$_2$ above 500 km:
\begin{equation}\tag{R80}
\hbox{CH} + \hbox{C$_2$H$_2$}  \longrightarrow  \hbox{$c$-C$_3$H$_2$} + \hbox{H} 
\end{equation}

Additional formation below 600 km is provided by
\begin{equation}\tag{R193}
  \hbox{H$_2$} + \hbox{$c$-C$_3$H}  \longrightarrow 
                                                       \hbox{$c$-C$_3$H$_2$}
                                                       + \hbox{H}
\end{equation}

Destruction above 600 km is by photodissociation forming $c$-C$_3$H,
$l$-C$_3$H, and C$_3$, and by reaction with CH$_3$ 
\begin{equation}\tag{R135}
  \hbox{$c$-C$_3$H$_2$} +\hbox{CH$_3$}  \longrightarrow 
  \hbox{C$_2$H$_2$} + \hbox{C$_2$H$_3$}
\end{equation}

Photodissociation and reaction with CH$_3$ continue to be important
below 600 km, but at these altitudes destruction also occurs via
\begin{equation}\tag{R24}
  \hbox{$c$-C$_3$H$_2$} + \hbox{H}   \xrightarrow{\text{M}} 
                                                                   \hbox{C$_3$H$_3$}  
\end{equation}
                                                                 and
\begin{equation}\tag{R180}
\hbox{$c$-C$_3$H$_2$} + \hbox{C$_2$H$_3$} \longrightarrow
\hbox{C$_3$H$_3$} + \hbox{C$_2$H$_2$}
\end{equation}

\begin{figure}
  \centering
  \includegraphics[width=0.8\linewidth]{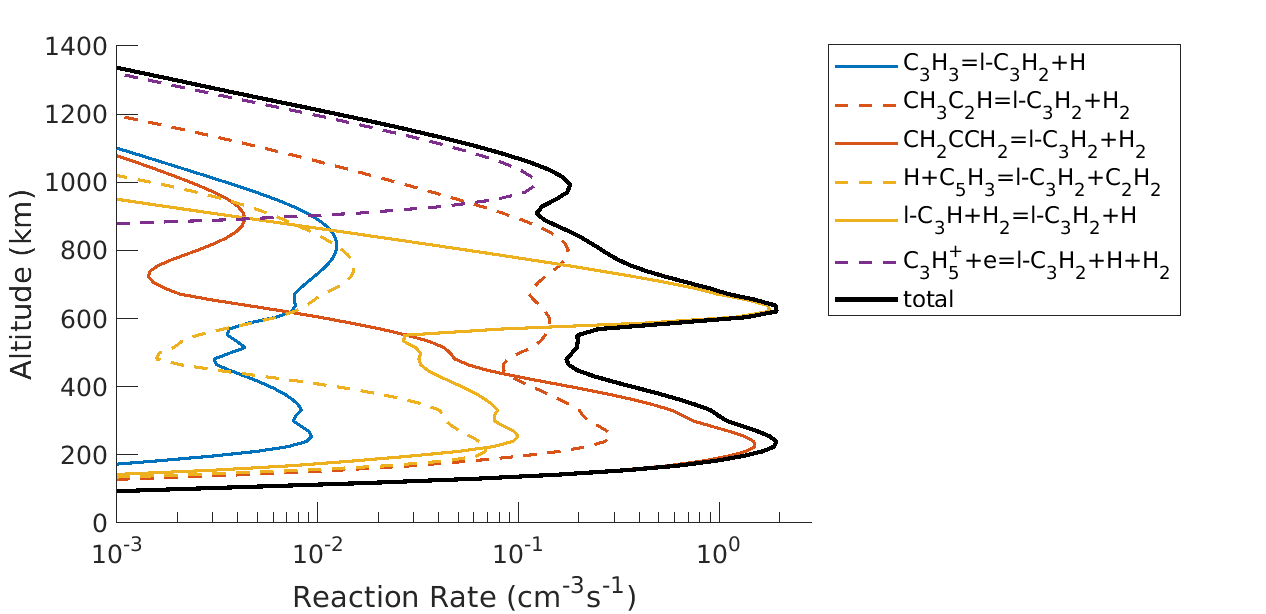}
  \includegraphics[width=0.8\linewidth]{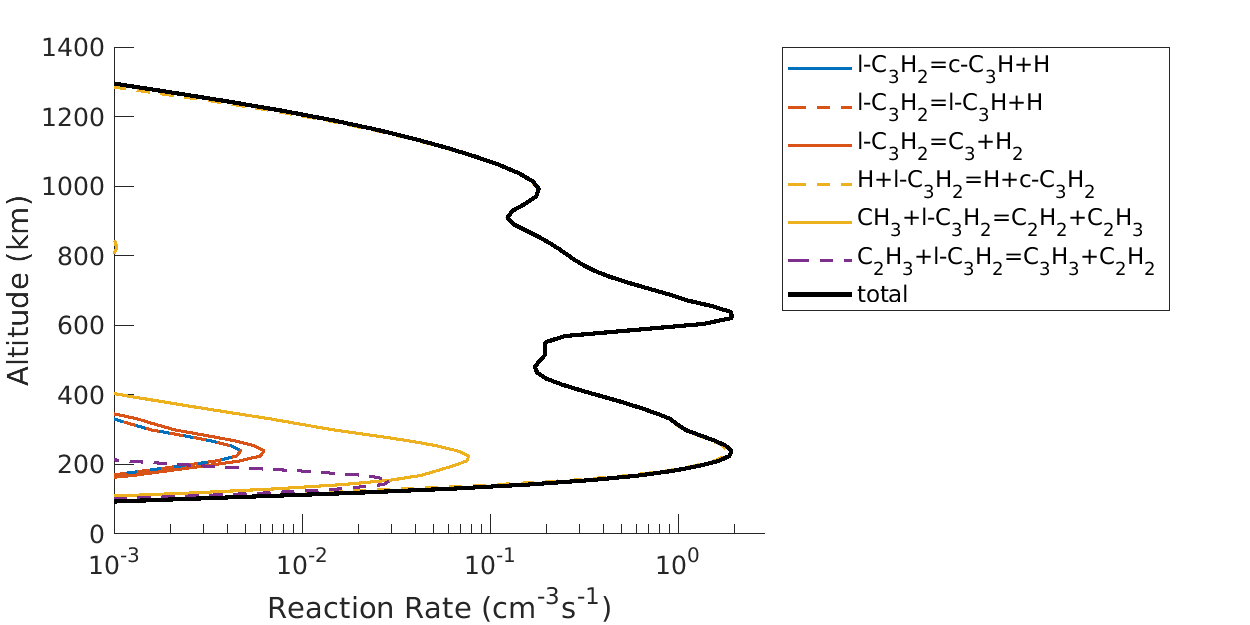}
  \caption{\label{fig:l-C3H2prod}Production (top) and
    destruction(bottom) rates for $l$-C$_3$H$_2$.  }
  \end{figure}

For $l$-C$_3$H$_2$ the formation processes are similar to $c$-C$_3$H$_2$,
with recombination of C$_3$H$_5^+$ (R1184) important at high altitudes,
along with photodissociation of CH$_2$CCH$_2$ and
CH$_3$C$_2$H (at all altitudes). Neutral-neutral reactions play a role below
1000 km (see Figure~\ref{fig:l-C3H2prod}).
\begin{equation}\tag{R54}
\hbox{H} + \hbox{C$_5$H$_3$}  \longrightarrow  \hbox{$l$-C$_3$H$_2$}
                                                  + \hbox{C$_2$H$_2$}
\end{equation}                                    
\begin{equation}\tag{R194}
\hbox{$l$-C$_3$H} + \hbox{H$_2$}  \longrightarrow  \hbox{$l$-C$_3$H$_2$}
                                                   + \hbox{H}
\end{equation}
Destruction is dominated by reaction with hydrogen atoms forming
$c$-C$_3$H$_2$ at all altitudes.  Below 400 km reaction with CH$_3$ and
C$_2$H$_3$ can also destroy $l$-C$_3$H$_2$ (Figure~\ref{fig:l-C3H2prod})
\begin{equation}\tag{R136}
\hbox{CH$_3$} + \hbox{$l$-C$_3$H$_2$}  \longrightarrow 
\hbox{C$_2$H$_2$} + \hbox{C$_2$H$_3$}  \hbox{~~~$<$ 400 km}
\end{equation}
\begin{equation}\tag{R181}
\hbox{C$_2$H$_3$} + \hbox{$l$-C$_3$H$_2$}  \longrightarrow 
\hbox{C$_3$H$_3$} + \hbox{C$_2$H$_2$} \hbox{~~~$<$ 200 km}
\end{equation}

\begin{figure}
  \centering
  \includegraphics[width=0.8\linewidth]{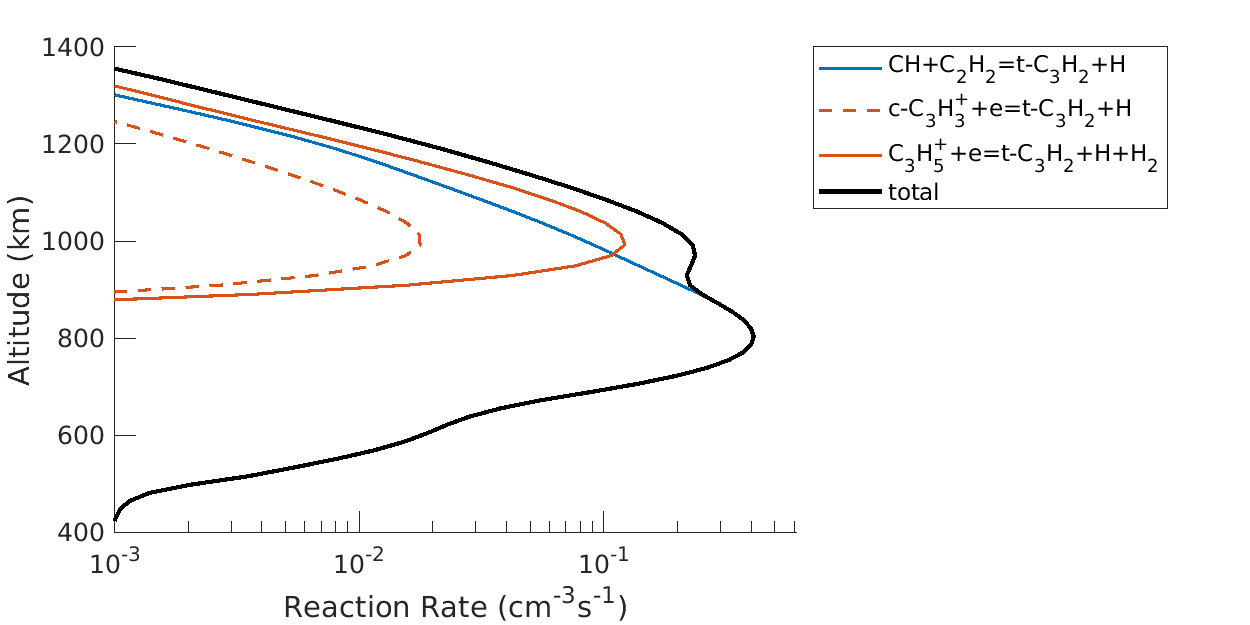}
  \includegraphics[width=0.8\linewidth]{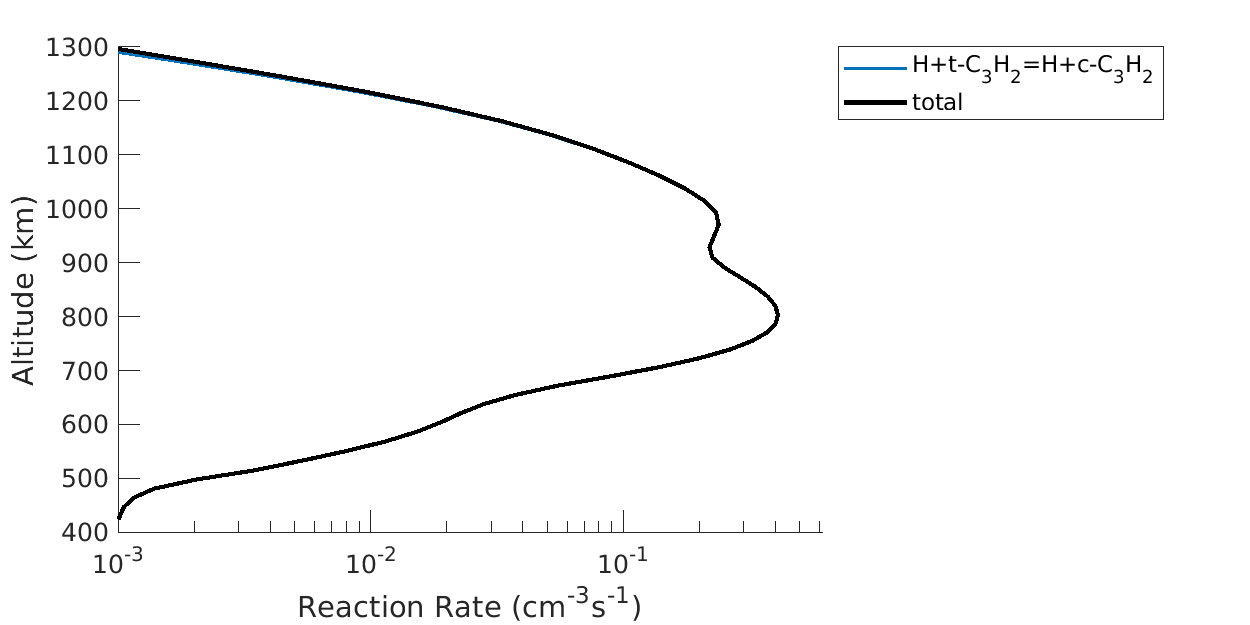}
  \caption{\label{fig:t-C3H2_prod}Production (top) and loss (bottom)
    rates for $t$-C$_3$H$_2$.}
  \end{figure}

The chemistry of $t$-C$_3$H$_2$ is different.  It forms by
\begin{equation}\tag{R83}
\hbox{CH} + \hbox{C$_2$H$_2$} \longrightarrow  \hbox{$t$-C$_3$H$_2$} + \hbox{H} \\
\end{equation}
with a major contribution from recombination of electrons with
$c$-C$_3$H$_3^+$ and $c$-C$_3$H$_5^+$ above 800 km.
Destruction at all altitudes is by reaction with H atoms to form
$c$-C$_3$H$_2$. The production and
loss rates are shown in Figure~\ref{fig:t-C3H2_prod}.

We will briefly discuss the peak structure seen in the abundance of
$c$-C$_3$H$_2$ and $l$-C$_3$H$_2$ at 600 km in Figure~\ref{fig:c3h2}.
A major contributor to this feature is the reaction of C$_3$ and
H$_2$ forming $l$-C$_3$H and $c$-C$_3$H which subsequently react with
H$_2$ to form $c$-C$_3$H$_2$ and
$l$-C$_3$H$_2$ at this
altitude.   The rate of this reaction is uncertain.  Here we use an estimated
value of \mbox{8.0 $\times$ 10$^{-12}$ e$^{(-1420/T)}$ cm$^{3}$s$^{-1}$} from \cite{kida}.  Changes to this
rate might reduce the size of the peak in the abundance of the C$_3$H$_2$ isomers.
In addition, there is a stagnant layer in the
vertical transport, as manifest in a minimum in the value for K$_{zz}$ in
Section~\ref{sec:models} (see also the red line in Fig. 4 of Li et al. 2014).  Both
of these factors contribute to the increase in $c$-C$_3$H$_2$ seen at this altitude.

\subsection{Chemistry of C$_3$H$_4$ - propadiene and methylacetylene}
\begin{figure}
    \centering
    \includegraphics[ width=0.48\linewidth]{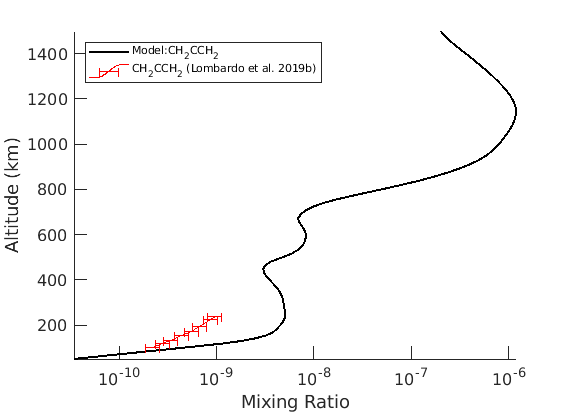}
    \includegraphics[width=0.48\linewidth]{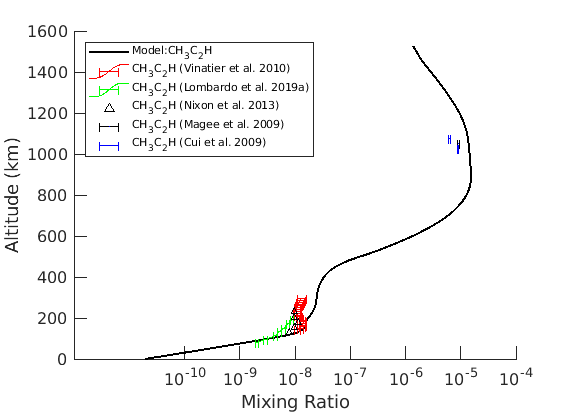}
    \caption{Mixing ratios of CH$_2$CCH$_2$ (left) and CH$_3$C$_2$H
      (right)  \label{fig:C3H4} }
\end{figure}

Both propadiene (CH$_2$CCH$_2$; also known as allene) and
methylactyelene (CH$_3$C$_2$H) have 
been observed in Titan's stratosphere \citep{lombardo19a,lombardo19b,
  nixon13,vinatier10}.  In addition, CH$_3$C$_2$H$_2$ was observed
using Cassini/UVIS (UltraViolet Imaging Spectrometer) \citep{cui09,magee09}. The models over predict the
abundances of these two molecules by a factor of a few
(Figure~\ref{fig:C3H4}) in the lower atmosphere. 
 
\begin{figure}
\centering
\includegraphics[width=0.8\linewidth]{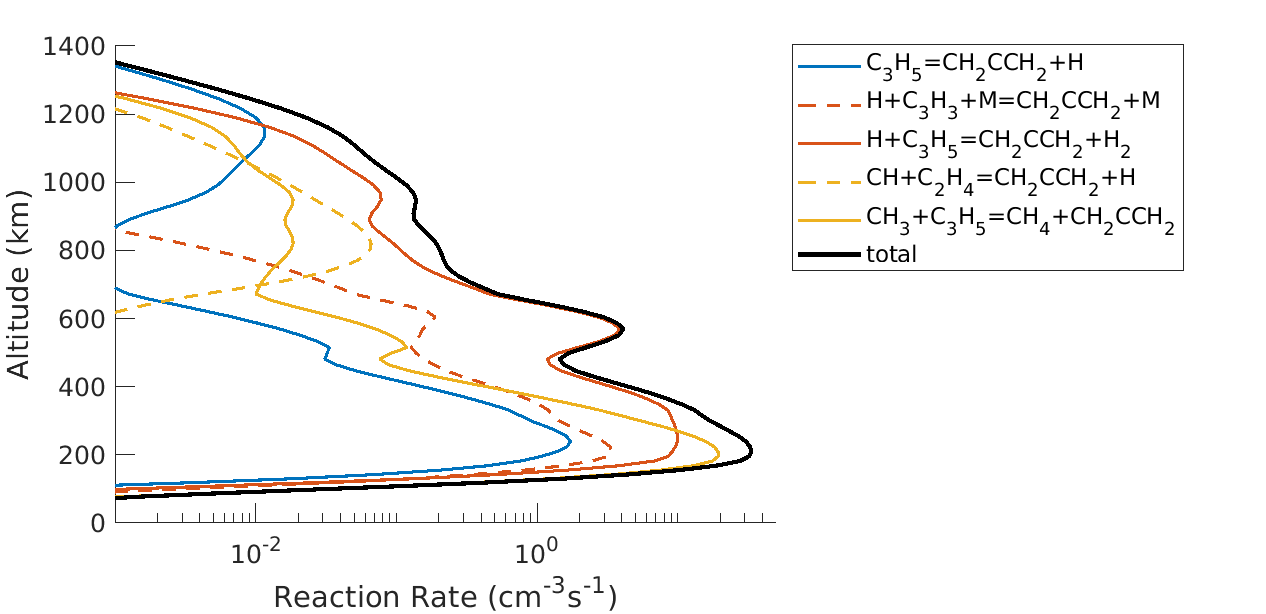}
\includegraphics[width=0.8\linewidth]{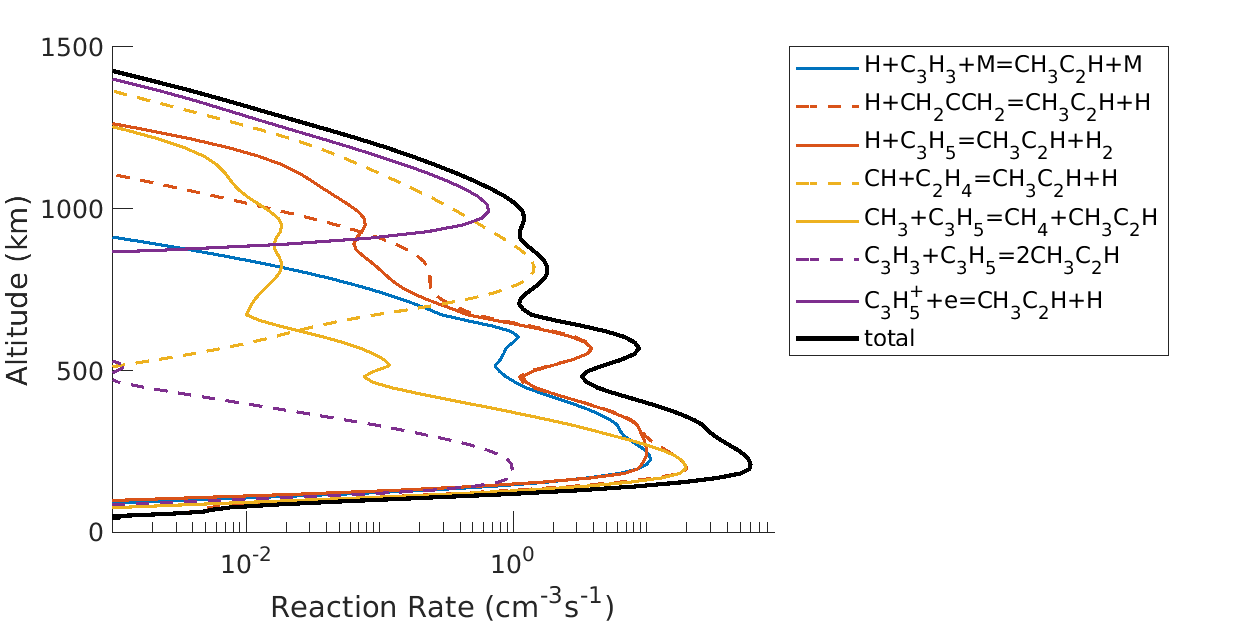}
\caption{\label{fig:c3h4_prod}Production rates of C$_3$H$_4$ isomers}
\end{figure}

Since there is only a small difference in enthalpy between CH$_3$C$_2$H and CH$_2$CCH$_2$ ($\sim$ 1 kcal mol$^{-1}$; Rogers and McIafferty 1995) their formation and loss rates are expected to be similar. Their main formation route is
\begin{eqnarray}
\hbox{H} + \hbox{C$_3$H$_5$} & \longrightarrow &
                                                 \left\{\begin{array}{l}
\hbox{CH$_2$CCH$_2$}  + \hbox{H$_2$}  \hbox{~~~50\%} \nonumber \hbox{~~~~~~~~~~(R32)}\\
 \hbox{CH$_3$C$_2$H} + \hbox{H$_2$} \hbox{~~~~50\%} \nonumber \hbox{~~~~~~~~~~(R31)}\\
\end{array}
\right.
\end{eqnarray}
with other important reactions being
\begin{eqnarray}
\hbox{CH$_3$} + \hbox{C$_3$H$_5$} & \longrightarrow &
\left\{ \begin{array}{l}
          \hbox{CH$_2$CCH$_2$} +  \hbox{CH$_4$}\nonumber  \hbox{~~~50\%}\hbox{~~~~~~~~~~(R140)}\\
          \hbox{CH$_3$C$_2$H}  + \hbox{CH$_4$}\nonumber  \hbox{~~~50\%}\hbox{~~~~~~~~~~(R139}\\
\end{array}
\right.
\end{eqnarray}
\begin{eqnarray}
\hbox{H} + \hbox{C$_3$H$_3$} & \xrightarrow{\text{M}} &
                                                        \left\{ \begin{array}{l}
\hbox{CH$_2$CCH$_2$} \nonumber  \hbox{~~~15\%}\hbox{~~~~~~~~~~(R25)}\\
\hbox{CH$_3$C$_2$H}\nonumber  \hbox{~~~85\%}\hbox{~~~~~~~~~~(R26)}\\
\end{array}
\right.
\end{eqnarray}

\begin{eqnarray}
\hbox{CH} + \hbox{C$_2$H$_4$} & \xrightarrow{\text{M}} & 
\left\{ \begin{array}{l}
\hbox{CH$_2$CCH$_2$} \nonumber \hbox{~~~50\%}\hbox{~~~~~~~~~~(R85)}\\
\hbox{CH$_3$C$_2$H}\nonumber \hbox{~~~50\%}\hbox{~~~~~~~~~~(R86)}\\
\end{array}
\right.
\end{eqnarray}
\cite{yap84} suggested isomerization as a way to convert CH$_2$CCH$_2$
into CH$_3$C$_2$H and this is the major route to CH$_3$C$_2$H
throughout much of the atmosphere:
\begin{equation}\tag{R29}
\hbox{H} + \hbox{CH$_2$CCH$_2$} \longrightarrow \hbox{CH$_3$C$_2$H}
+ \hbox{H}
\end{equation}
Additionally, CH$_3$C$_2$H is also formed above 800 km by
electron  recombination of C$_3$H$_5^+$ and below 500 km by
\begin{equation}\tag{R199}
\hbox{C$_3$H$_3$} + \hbox{C$_3$H$_5$} \longrightarrow
\hbox{2 CH$_3$C$_2$H} 
\end{equation}
CH$_2$CCH$_2$ is also formed by photodissociation of C$_3$H$_5$. 
The formation reactions of CH$_3$C$_2$H
and CH$_2$CCH$_2$ are summarized in Figure~\ref{fig:c3h4_prod}.

\begin{figure}
\centering
\includegraphics[width=0.8\linewidth]{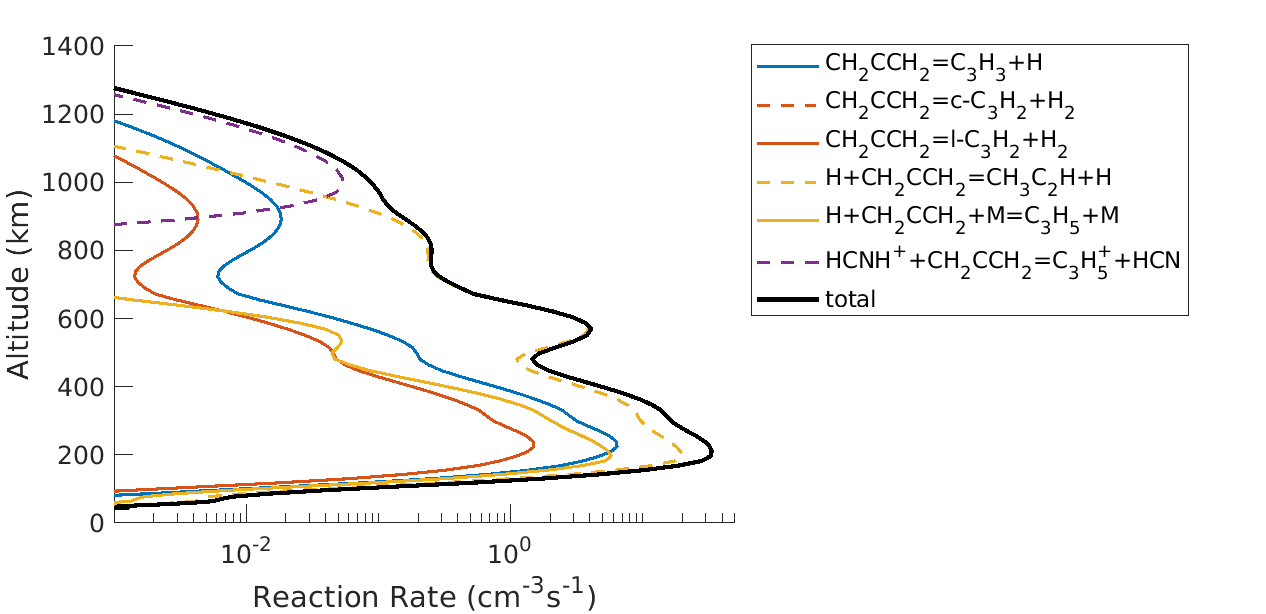}
\includegraphics[width=0.8\linewidth]{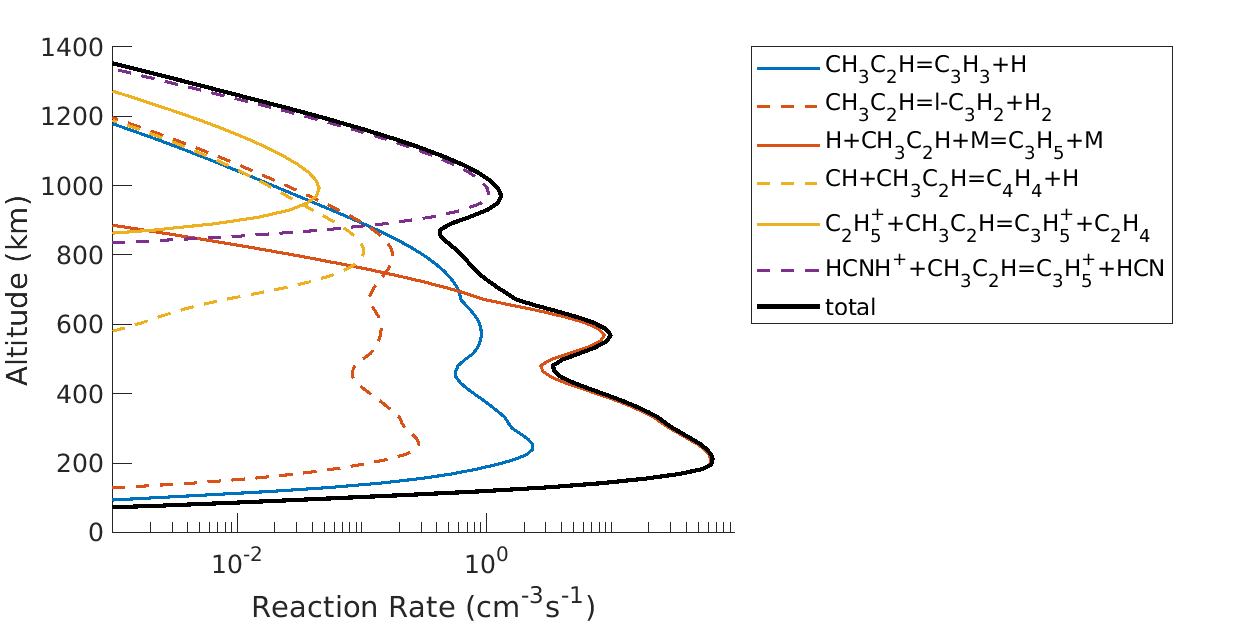}
\caption{\label{fig:c3h4_loss}Loss rates of C$_3$H$_4$ isomers}
\end{figure}

Destruction of both molecules (Figure~\ref{fig:c3h4_loss}) is by
photodissociation at high altitudes, forming $c$-C$_3$H$_2$ and
$l$-C$_3$H$_2$.  In this region both molecules are
also destroyed by reaction with HCNH$^+$
\begin{equation}
\left.
\begin{array}{l}
\hbox{HCNH$^+$} + \hbox{CH$_3$C$_2$H} \\
\hbox{HCNH$^+$} + \hbox{CH$_2$CCH$_2$} \end{array} \right\}  \longrightarrow  \hbox{HCN} +
\hbox{C$_3$H$_5^+$}  \begin{array}{l} \hbox{~~~~~~~~~~~~(R1062)}\\ \hbox{~~~~~~~~~~~~(R1063)} \end{array}
\nonumber
\end{equation}

Below 600 km both molecules are destroyed by reaction with H atoms, with CH$_2$CCH$_2$  undergoing
isomerization, and (in a three-body reaction) both molecules forming C$_3$H$_5$:
\begin{equation}\tag{R29}
\hbox{CH$_2$CCH$_2$} + \hbox{H}  \longrightarrow 
                                                   \hbox{CH$_3$CCH$_2$}
                                                   +
                                                   \hbox{H} \label{eq:isom}
\end{equation}
\begin{equation}
\left.
\begin{array}{l}
\hbox{CH$_3$C$_2$H} + \hbox{H} \\
\hbox{CH$_2$CCH$_2$} + \hbox{H} \end{array} \right\} \xrightarrow{\text{M}} 
\hbox{C$_3$H$_5$} 
\begin{array}{l}
\hbox{~~~~~~~~~~~~(R28)} \\ \hbox{~~~~~~~~~~~~(R30)}\end{array}
\nonumber
\end{equation}

\subsection{\label{sec:c3h6}C$_3$H$_6$}

\begin{figure}
\centering
\includegraphics[width=0.48\linewidth]{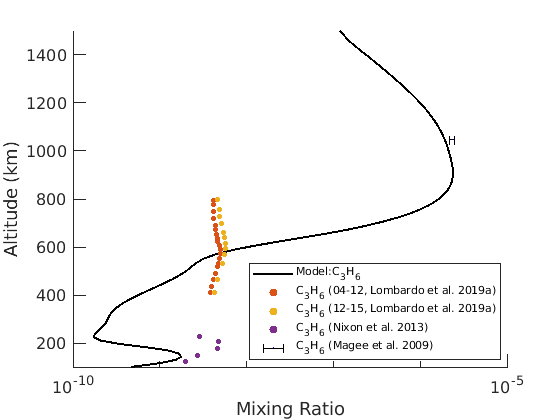}
\caption{\label{fig:c3h6}Model predictions (solid black line) and
  observations (shown by symbols) for abundance of
  C$_3$H$_6$. }
\end{figure}

Propylene (C$_3$H$_6$) has been observed by \cite{lombardo19a},
\cite{nixon13} and \cite{magee09}.  The model produces fractional
abundances in good agreement wih the observations both in the lower
atmosphere (below 400 km) and at $\sim$ 1000 km (Figure~\ref{fig:c3h6}).  

\begin{figure}
\centering
    \includegraphics[width=0.8\linewidth]{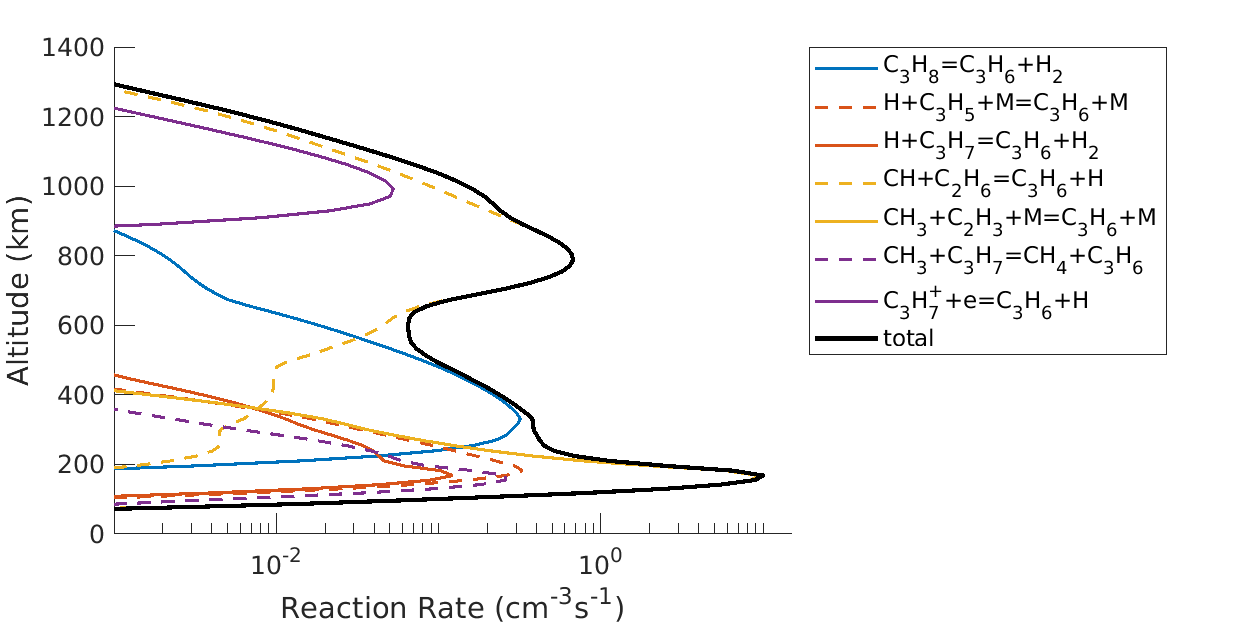}
    \includegraphics[width=0.8\linewidth]{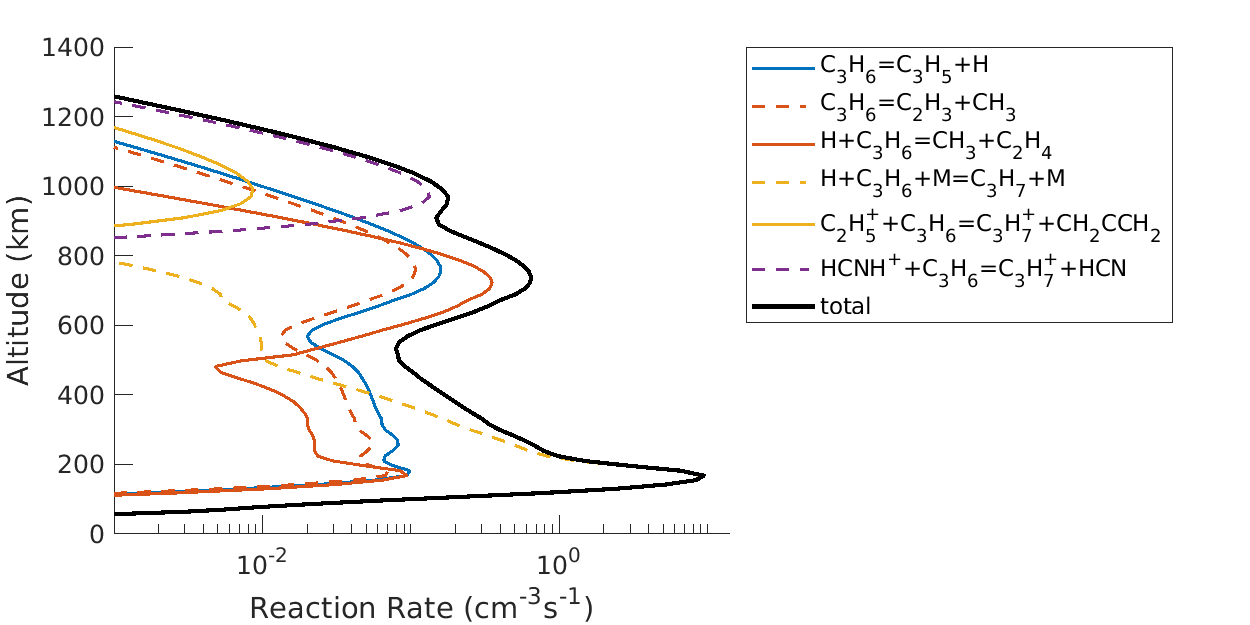}
\caption{\label{fig:c3h6_prod}The main production and loss reactions
  for propylene. }
\end{figure}

The main production and destruction reactions are shown in
Figure~\ref{fig:c3h6_prod}.  Below 400 km production is dominated by 
\begin{equation}\tag{R132}
\hbox{CH$_3$} + \hbox{C$_2$H$_3$} \xrightarrow{\text{M}}
\hbox{C$_3$H$_6$} + \hbox{CH$_4$}
\end{equation}
\begin{equation}\tag{R34}
  \hbox{H} + \hbox{C$_3$H$_5$} \xrightarrow{\text{M}}
  \hbox{C$_3$H$_6$}
  \end{equation}
\begin{equation}\tag{R144}
\hbox{CH$_3$} + \hbox{C$_3$H$_7$} \longrightarrow \hbox{C$_3$H$_6$} +
\hbox{CH$_4$}
\end{equation}
\begin{equation}\tag{R38}
\hbox{H} + \hbox{C$_3$H$_7$} \longrightarrow \hbox{C$_3$H$_6$} +
\hbox{H$_2$}
\end{equation}
Between 400 and 600 km the main production route is by
photodissociation of C$_3$H$_8$, and above this reactions with CH and
C$_3$H$_7^+$ dominate:
\begin{equation}\tag{R88}
\hbox{CH} + \hbox{C$_2$H$_6$}  \longrightarrow  \hbox{C$_3$H$_6$} +
                                                  \hbox{H}
\end{equation}
\begin{equation}\tag{R1203}
\hbox{C$_3$H$_7^+$} + e  \longrightarrow  \hbox{C$_3$H$_6$} +
                                           \hbox{H}
\end{equation}

At high altitudes ($>$ 800km) destruction is by reaction with HCNH$^+$
forming HCN, and by 
\begin{equation}\tag{R747}
\hbox{C$_2$H$_5^+$} + \hbox{C$_3$H$_6$} \longrightarrow \hbox{C$_3$H$_7^+$}
+ \hbox{CH$_2$CCH$_2$}
\end{equation}

Photodissociation forming C$_2$H$_2$, C$_2$H$_3$ and C$_3$H$_5$ is
important at most altitudes.  Below 400 km the main loss processes are
\begin{equation}\tag{R37}
\hbox{C$_3$H$_6$} + \hbox{H} \xrightarrow{\text{M}} \hbox{C$_3$H$_7$}
\end{equation}.
\begin{equation}\tag{R36}
  \hbox{C$_3$H$_6$} + \hbox{H} \longrightarrow \hbox{CH$_3$} +
  \hbox{C$_2$H$_4$}
\end{equation}

\subsection{\label{sec:c3h8}C$_3$H$_8$}

\begin{figure}
\centering
\includegraphics[width=0.8\linewidth]{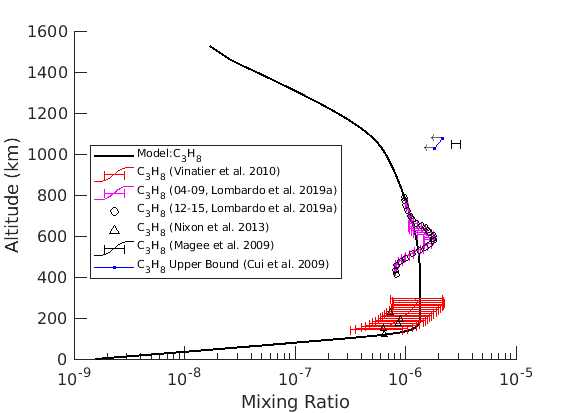}
\caption{\label{fig:c3h8}Abundance of propane.  Solid line is model, symbols are the
  observations from \cite{vinatier10} (red), \cite{lombardo19a} (pink
  shows data from 2004-2009, and open circles data from 2012-2015), and
\cite{nixon13} (open triangles). Also shown are INMS data from Cui et
al. (2009; upper limits in blue), and Magee et al. (2009; black).}
\end{figure}

\begin{figure}
\centering
\includegraphics[width=0.8\linewidth]{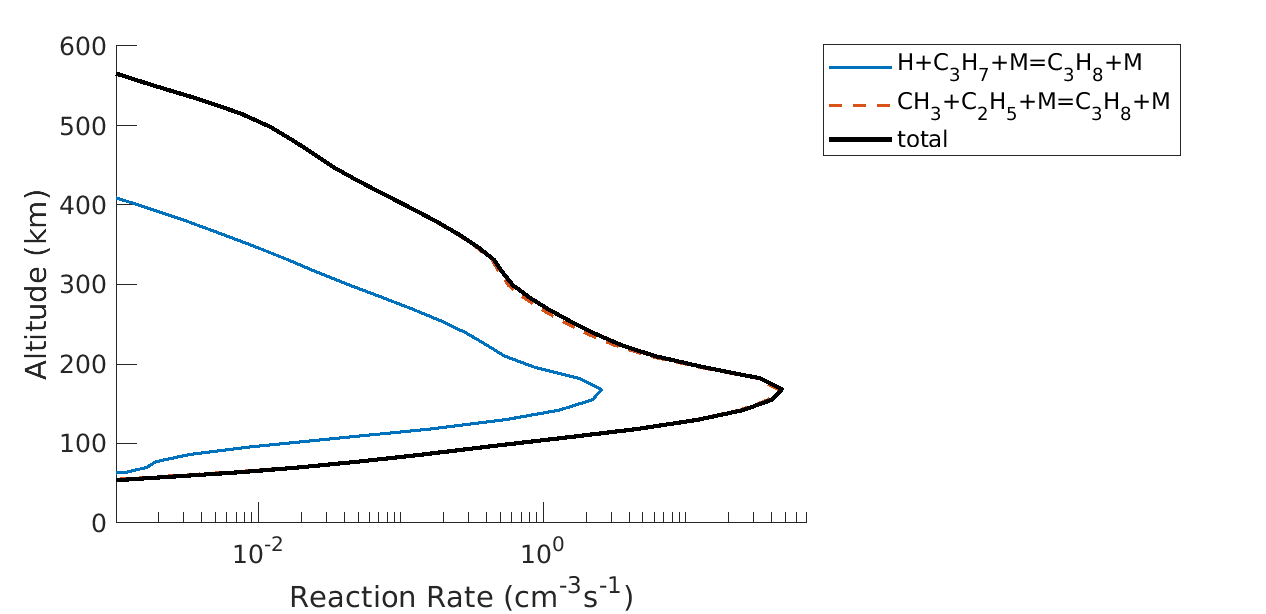}
\includegraphics[width=0.8\linewidth]{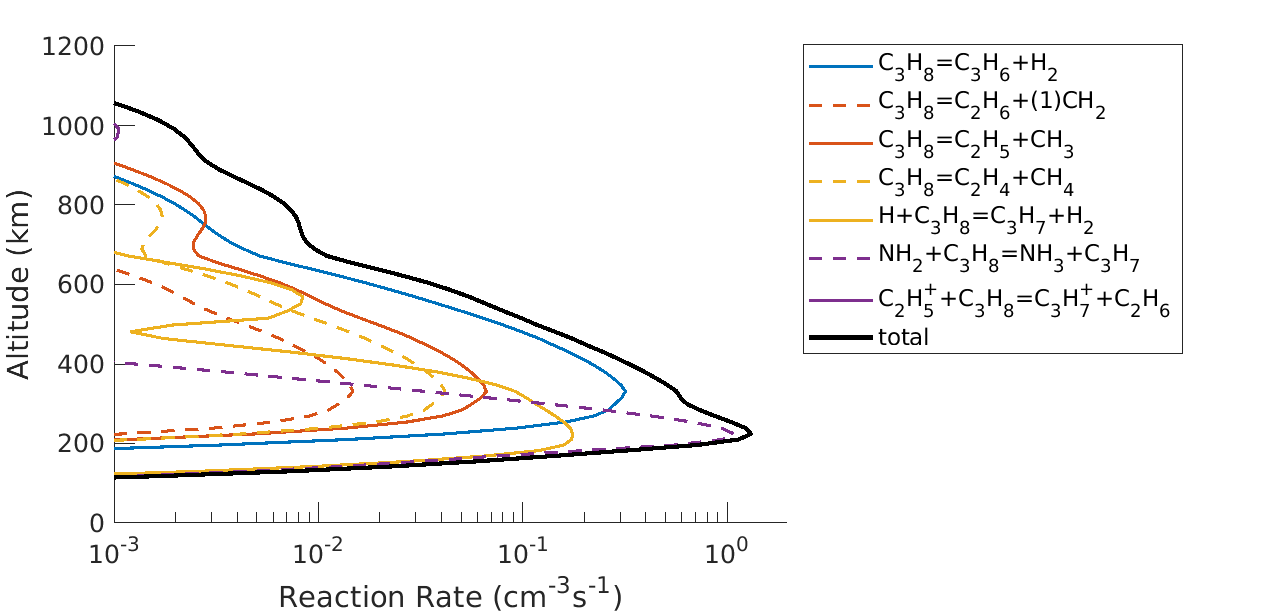}
\caption{\label{fig:c3h8_prod}The main production and loss reactions
  for propane. }
\end{figure}

The predicted abundance distribution  of C$_3$H$_8$ is in excellent
agreement with the observations of \cite{vinatier10, lombardo19a} and
\cite{nixon13} (Figure~\ref{fig:c3h8}).  In the upper atmosphere the
models are in agreement with the INMS upper limits \citep{cui09,magee09}.
The main production and loss reactions of propane are shown in Figure~\ref{fig:c3h8_prod}.
Production is by 
\begin{equation}\tag{R134}
\hbox{CH$_3$} + \hbox{C$_2$H$_5$} \xrightarrow{\text{M}} \hbox{C$_3$H$_8$}
\end{equation}
and 
\begin{equation}\tag{R40}
\hbox{H} + \hbox{C$_3$H$_7$} \xrightarrow{\text{M}} 
                                                        \hbox{C$_3$H$_8$}
\end{equation}
while destruction is mainly by photodissociation
\begin{eqnarray}
\hbox{C$_3$H$_8$} & \xrightarrow{\text{h$\nu$}} & \left\{
   \begin{array}{l}
\hbox{C$_2$H$_4$} + \hbox{CH$_4$} \\
\hbox{C$_2$H$_5$} + \hbox{CH$_3$}\\
\hbox{C$_2$H$_6$} + \hbox{CH$_2$}\\
\hbox{C$_3$H$_6$} + \hbox{H$_2$}\\
   \end{array}
  \right.
  \nonumber
\end{eqnarray}
with an additional contribution from reaction with H atoms forming
C$_3$H$_7$.
Below 400 km C$_3$H$_8$ reacts with NH$_2$ to form NH$_3$ and
C$_3$H$_7$ (R291).

\subsection{C$_3$H$_3$ and C$_6$H$_6$}
\begin{figure}[!h]
    \centering
    \includegraphics[width=0.8\linewidth]{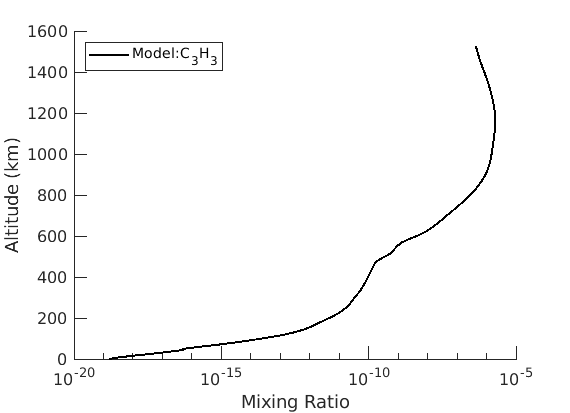}
    \caption{Mixing ratio of C$_3$H$_3$.    \label{fig:C3H3}}
\end{figure}

\begin{figure}[!h]
\centering
    \includegraphics[ width=0.8\linewidth]{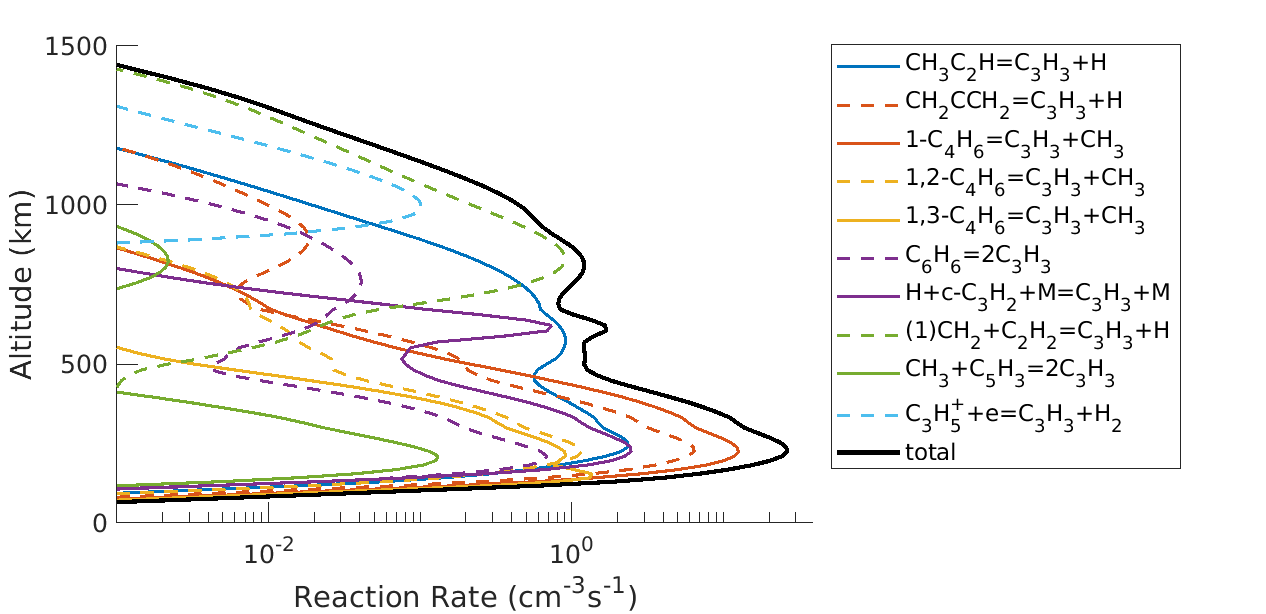}
    \includegraphics[width=0.8\linewidth]{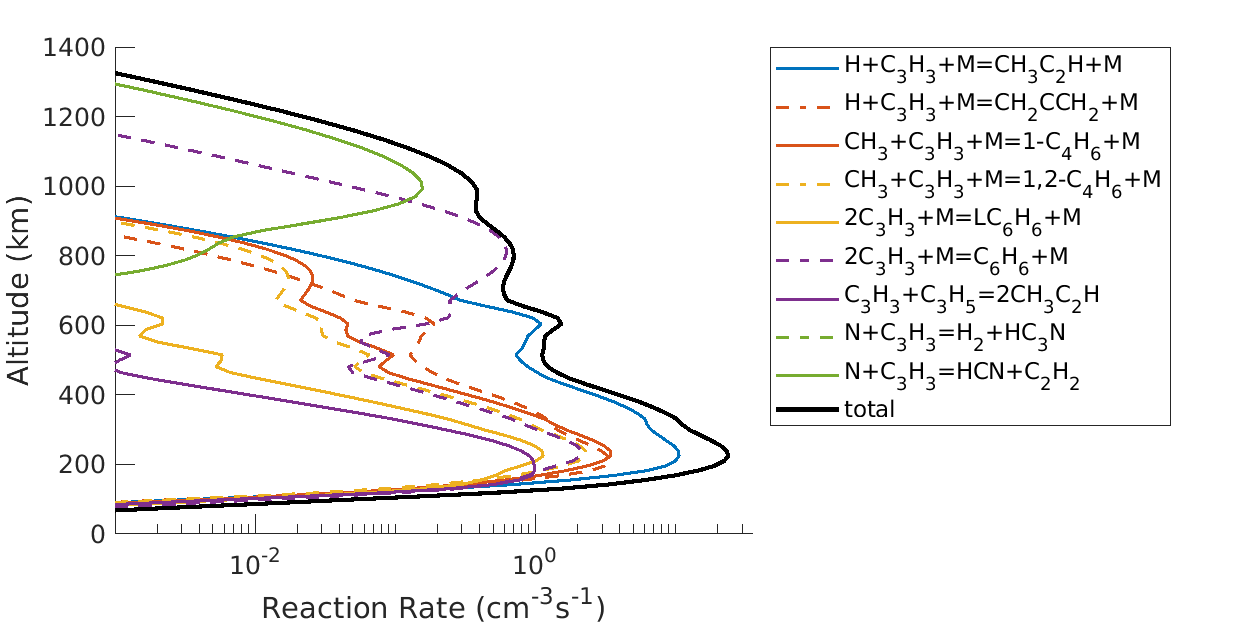}
\caption{\label{fig:c3h3_prod}Main production and loss processes for
  the propargyl radical. }
\end{figure}

Although not observed the propargyl radical plays a crucial role in
the formation of benzene
(Figure~\ref{fig:C3H3}).   Its peak abundance is 2.5 $\times$ 10$^{-6}$
at 1050 km and it is produced by the  photodissociation of CH$_2$CCH$_2$, CH$_3$C$_2$H,
1-C$_4$H$_6$, 1,2-C$_4$H$_6$ and 1,3-C$_4$H$_6$ at all
altitudes  (Figure~\ref{fig:c3h3_prod}). These are supplemented by the recombination of C$_3$H$_5^+$
with electrons at altitudes above 1000 km as well as
\begin{equation}\tag{R104}
\hbox{$^1$CH$_2$} + \hbox{C$_2$H$_2$}  \longrightarrow 
\hbox{C$_3$H$_3$} + \hbox{H}
\end{equation}
Below 800 km it can also
form by reaction of H and $c$-C$_3$H$_2$.
\begin{equation}\tag{R24}
\hbox{H} + \hbox{$c$-C$_3$H$_2$} \xrightarrow{\text{M}}
\hbox{C$_3$H$_3$}
\end{equation}

Destruction is by
\begin{equation}\tag{R198}
  \hbox{2 C$_3$H$_3$}   \xrightarrow{\text{M}}  \hbox{C$_6$H$_6$}
                                                 \hbox{~~~ $<$ 1200
                                                 km}
\end{equation}
\begin{equation}
\hbox{H} + \hbox{C$_3$H$_3$}   \xrightarrow{\text{M}}  \left\{
\begin{array}{ll}
\hbox{CH$_3$C$_2$H}  & \hbox{~~~ $<$ 600 km} \hbox{~~~~~~~~~~(R31)}\\
\hbox{CH$_2$CCH$_2$}  &\hbox{~~~ $<$ 600 km} \hbox{~~~~~~~~~~(R32)} \nonumber\\
\end{array}
\right.
\end{equation}

Above 1000 km reactions with N dominate
\begin{equation}
\hbox{N} + \hbox{C$_3$H$_3$} \longrightarrow  \left\{
\begin{array}{l}
 \hbox{HCN} + \hbox{C$_2$H$_2$} \hbox{~~~~~~~~~~(R237)}\\
\hbox{HC$_3$N} + \hbox{H$_2$} \hbox{~~~~~~~~~~(R236)} \nonumber\\
\end{array}
\right.
\end{equation}

Benzene is formed from neutral-neutral
reactions involving C$_3$H$_3$, and this is the main formation route
throughout much of the atmosphere below 1000 km.  

\begin{eqnarray}
 2 \hbox{C$_3$H$_3$} &   \xrightarrow{\text{M}} & \left\{
\begin{array}{lll}
\hbox{$l$-C$_6$H$_6$} & \xrightarrow{\text{H}} & \hbox{C$_6$H$_6$} \hbox{~~~~~~~~~~~~~~~~(R197 and R222)}\\
                                                     \hbox{C$_6$H$_6$} 
                    & &\hbox{~~~~~~~~~~~~~~~~~~~~~~~~(R198)} \nonumber \\
\end{array}
\right.
\end{eqnarray}

\begin{figure}
  \centering
  \includegraphics[width=0.6\linewidth]{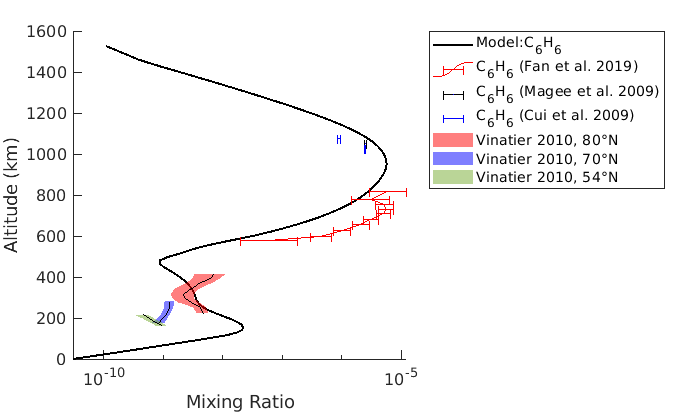}
    \caption{\label{fig:C6H6}Model predictions of benzene mixing ratio in Titan's
    atmosphere (solid line). Open circles show the observed abundance
    derived by \cite{fan19} from Cassini/UVIS data.  In the
    stratosphere observations are from \cite{vinatier10}. }
\end{figure}

The abundance of benzene therefore provides a test of the model predictions of
the C$_3$H$_3$ radical.  Figure~\ref{fig:C6H6} shows the predicted benzene
abundance with the measured abundances retrieved from UVIS
observations \citep{fan19}. The model provides a good fit to the
observations at all altitudes, suggesting that our model also
correctly describes the chemistry of the propargyl radical.

Above 1000 km ion chemistry is important for benzene formation through the recombination of
C$_6$H$_7^+$ with electrons.
\begin{equation}\tag{R1230}
  \hbox{C$_6$H$_7^+$} + \hbox{$e^-$} \longrightarrow \hbox{C$_6$H$_6$}
  + \hbox{H}  \nonumber
\end{equation}

C$_6$H$_7^+$ forms from
\begin{equation}\tag{R848}
\hbox{C$_4$H$_5^+$} + \hbox{C$_2$H$_4$}  \longrightarrow 
                                                        \hbox{C$_6$H$_7^+$}
                                                        + \hbox{H$_2$} \\
\end{equation}
\begin{equation}\tag{R825}
\hbox{C$_3$H$_5^+$} + \hbox{CH$_3$C$_2$H} \longrightarrow 
                                                          \hbox{C$_6$H$_7^+$}
                                                          + \hbox{H$_2$} \nonumber
\end{equation} 

Destruction of benzene is by photodissociation forming C$_6$H$_5$, C$_2$H$_2$ and
C$_3$H$_3$.

{The major reactions involved in the benzene chemistry predicted by this model are consistent with 
 the models of \cite{vuitton08} and \cite{lavvas08a,lavvas08b}.}

\subsection{Ions}
\begin{figure}
\centering
\includegraphics[width=0.9\linewidth]{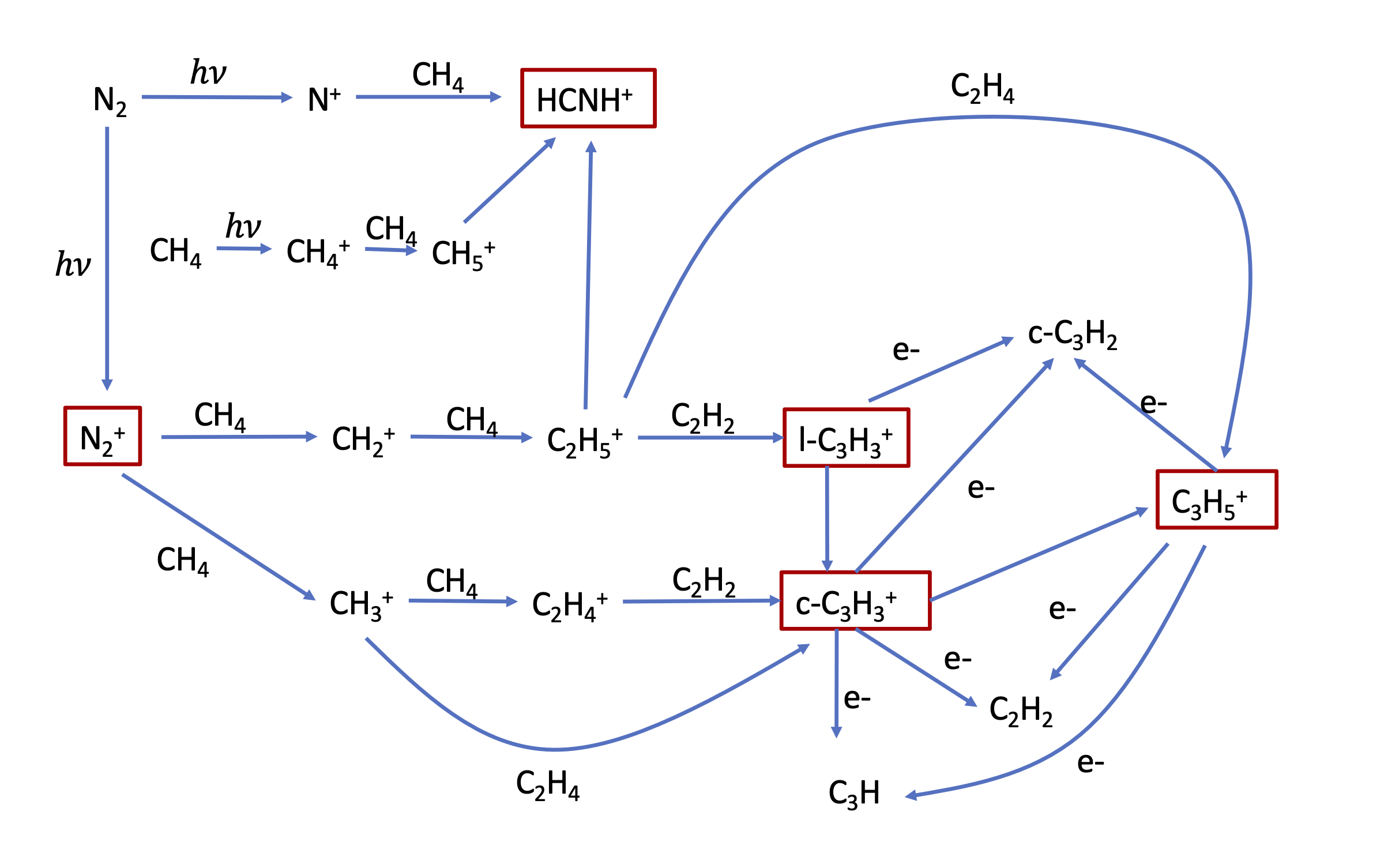}
\caption{\label{fig:ion_prod}Ion chemistry in the atmosphere of Titan
  showing the formation of the main ions involved in the C$_3$H$_n$
  chemistry (indicated by red boxes).  }
\end{figure}

The ion chemistry is initial by photoionization of N$_2$ and CH$_4$
forming N$_2^+$, N$^+$ and CH$_4^+$. Reactions of these ions drive the
ion chemistry.  The formation of ions involved in the C$_3$H$_n$
chemistry is shown in Figure~\ref{fig:ion_prod}.  

\begin{figure}
\centering
\includegraphics[width=0.48\linewidth]{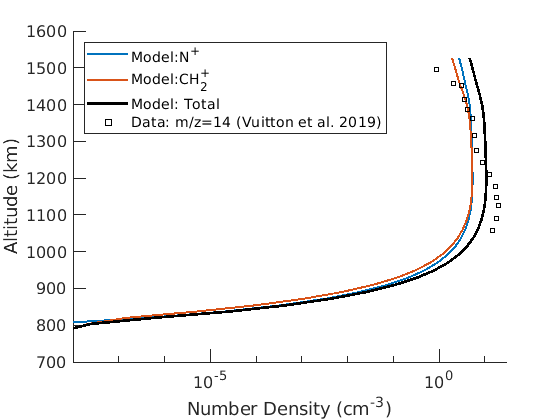}
\includegraphics[width=0.48\linewidth]{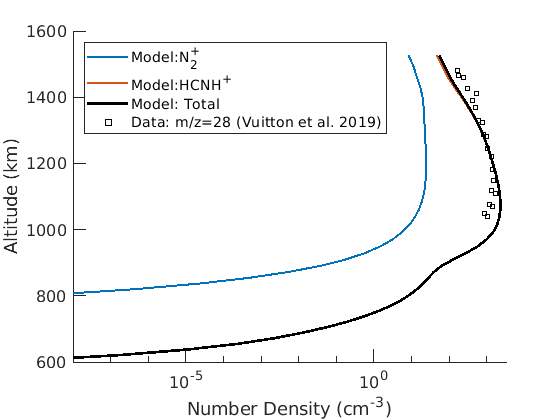}
\includegraphics[width=0.48\linewidth]{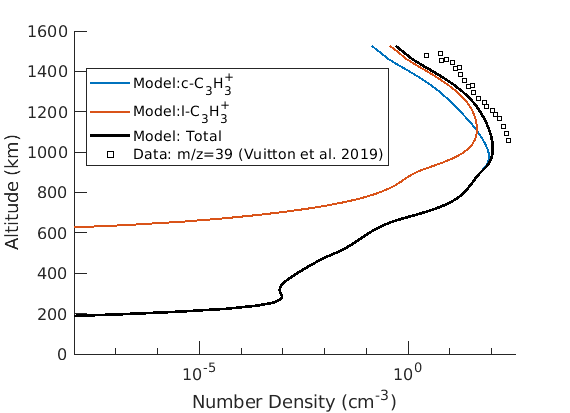}
\includegraphics[width=0.48\linewidth]{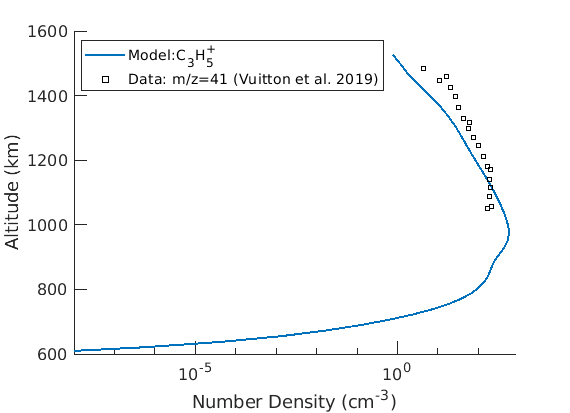}
\caption{\label{fig:ion_obs}Predicted number densities of ions with m/z
  of {\it(Top left:} 14, {\it {Top right:}} 28, {\it {Bottom left:}}
  39 and {\it {Bottom right:}} 41. Each panel shows the abundance
  of ions with the given m/z as well as the total abundance of all
  ions for that m/z (solid lines).  Open circles indicate the INMS
  observations.}
\end{figure}

Ion observations are available in the upper atmosphere Cassini/INMS
\citep{mandt12}. This provides the total number density of ions of a particular
mass/charge ratio, but cannot distinguish between ions of a given
mass.
Figure~\ref{fig:ion_obs} shows the predicted and
observed number densities of ions with m/z = 14 (N$^+$ and CH$_2^+$), 28
(N$_2^+$ , HCNH$^+$ and C$_2$H$_4^+$), 39 ($l$-C$_3$H$_3^+$, and
$c$-C$_3$H$_3^+$) and 41 (C$_3$H$_5^+$).  The observations cannot
distinguish between different ions of the same mass, but the figure
shows that the total abundance of ions with each mass is in good
agreement with the observations.

\section{\label{sec:disc}Discussion and conclusions}

It is a tribute to the recent advances in remote sensing that we are
now able to detect species such as cyclopropenylidene with mixing
ratios on the order of parts per billion by volume (10$^{-9}$) in the
mesosphere of Titan, or column densities on the order of 10$^{12}$
molecules cm$^{-2}$. Multiple in-situ and remote sensing techniques
contribute to the detection of benzene. At the top of the atmosphere,
in the thermosphere and ionosphere, we have measurements from
Cassini-Ion and Neutral Mass Spectrometer (INMS) \citep{waite07}.  In the mesosphere, we have Cassini-UVIS
remote sensing of benzene \citep{fan19}.  In the stratosphere we have
mid-infrared data from Cassini Composite Infrared Spectrometer (CIRS)
\citep{flasar05,coustenis07,vinatier10}. Finally, we have
Cassini-Huygens Gas Chromatograph Mass Spectrometer (GCMS)
measurements at the surface from the probe \citep{niemann05}. 

Our model provides excellent agreement with the available data for the abundances of the C$_3$ hydrocarbons. It
reproduces the column density of $c$-C$_3$H$_2$ to within a factor of
3, as well as its extracted distribution with altitude.  The
abundances of 
C$_3$H$_6$ and C$_3$H$_8$  are also well-matched, although the models
over predict the abundances of  propadiene and
methylacetylene in the lower atmosphere by factors of a few.
 The abundance of benzene is also in good agreement
with the abundances derived from the UVIS observations.  Since benzene
forms mainly from reactions of C$_3$H$_3$ this suggests that the model
predictions of this (unobserved) precursor molecule are also reasonable.
The possibility that the detection of benzene at the surface of Titan
may require a new production mechanism that is different from the ion
chemistry and C$_3$H$_3$ chemistry studied in this paper was suggested
by \citep{zhou10}. These authors propose a solid-phase chemistry
driven by cosmic rays, which effectively convert acetylene to benzene
\begin{equation}
  \hbox{3 C$_2$H$_2$} \longrightarrow \hbox{C$_6$H$_6$}
\end{equation}
However we notice from Figure~\ref{fig:C6H6} that the standard gas
phase mechanism is capable of generating a benzene mixing ratio of
10$^{-9}$. A definitive value of benzene from the GCMS experiment is
needed to confirm whether this additional cosmic ray driven mechanism is justified.

The successful modeling of benzene opens up the exciting possibility
of forming polycyclic aromatic hydrocarbons (PAHs) in planetary
atmospheres. This possibility was first suggested by
\cite{wong00,wong03} for the polar region of Jupiter. This type of
chemistry could eventually lead to the production of dark aerosols
\citep{zhang13}. Recent advances in experimental and theoretical
studies of the formation of multi-ringed species \citep{yang21}
confirmed our earlier speculations, opening up new avenues of
investigation for the planets in our Solar System and early Earth
\citep{berry19}. 

Ringed species are of great astrobiological significance 
\citep[see, e.g.,][]{sebree18}.  Although only two ringed molecules
($c$-C$_3$H$_2$ and benzene) have so far been detected in Titan, several
have been observed in the interstellar medium (ISM) \citep[see ][]{mcguire18}:
$c$-C$_3$H (propynylidyne), $c$-C$_3$H$_2$ (cyclopropenylidene), $c$-H$_2$C$_3$O
(cyclopropenone), $c$-C$_2$H$_4$O (ethylene oxide), C$_6$H$_6$ (benzene), $c$-C$_6$H$_5$CN
(benzonitrile). More recently, two isomers of cyanonaphthalene
(C$_{11}$H$_7$N),1-cyanonaphthalene, 2-cyanonaphthalene, have been
discovered \cite{mcguire21}. They are the first two-ring hydrocarbons
to be detected. The study of ring molecules in the ISM provides a
useful source for expanding our knowledge of planetary
atmospheres. The ISM is primarily driven by ion chemistry, which is
also important in the upper atmosphere of Titan.  If similar processes
can occur in Titan then it is likely that more ring molecules await
discovery and potentially more complex molecules of astrobiological significance.

\begin{acknowledgements}
\noindent  © 2022. All rights reserved. The research was carried out at the Jet Propulsion Laboratory,
California Institute of Technology, under a contract with the National
Aeronautics and Space Administration (80NM0018D0004).  It was
supported by the NASA Astrobiology Institute. We thank Run-Lie Shia
for help with modifying the KINETICS code used in our calculations.
\end{acknowledgements}

\clearpage
\newpage

\appendix
\section{\label{app:react}Reaction List}

\setlength\LTleft{0pt}
\setlength\LTright{0pt}

\begin{longtable}[l]{llll}
\caption{\label{tab:react}List of photochemical reactions.}\\
Number & Reaction & Rate  & Reference\\
\hline
 \rownumber & 2H       + M       = H$_2$      + M                        &     $k_0$ = 2.70 $\times$ 10$^{-31}$  T$^{-0.6}$    & \cite{moses05}\\ 
& & $k_{\infty}$ = 1.0 $\times$ 10$^{-11}$   & \cite{moses05}\\ \rownumber &
 H       + CH      = C       + H$_2$                           & 1.30 $\times$ 10$^{-10}$   exp(-80/T) & \cite{harding93}\\ \rownumber & 
 H       + $^{1}$CH$_2$  = CH      + H$_2$                           & 2.7 $\times$ 10$^{-10}$      &  \cite{kida} \\ \rownumber & 
  H       + $^3$CH$_2$  = CH      + H$_2$                           & 2.66 $\times$ 10$^{-10}$  & \cite{moses05} \\ \rownumber & 
  H       + $^3$CH$_2$  + M       = CH$_3$     + M                  & k$_0$ = 9.00 $\times$ 10$^{-32}$   exp(-550./T)  & \cite{moses11}\\
\end{longtable}


\end{document}